# 3T MR-Guided Brachytherapy for Gynecologic Malignancies

**Tina Kapur, PhD**[1], **Jan Egger, PhD**[1], **Antonio Damato, PhD**[2], **Ehud J. Schmidt, PhD**[1], and **Akila N. Viswanathan, MD, MPH**[2]

[1]Department of Radiology, Brigham and Women's Hospital Harvard Medical School, Boston, MA, USA

[2]Department of Radiation Oncology, Brigham and Women's Hospital Harvard Medical School, Boston, MA, USA

## Abstract

Gynecologic malignancies are a leading cause of death in women worldwide. Standard treatment for many primary and recurrent gynecologic cancer cases includes a combination of external beam radiation, followed by brachytherapy. Magnetic Resonance Imaging (MRI) is benefitial in diagnostic evaluation, in mapping the tumor location to tailor radiation dose, and in monitoring the tumor response to treatment. Initial studies of MR-guidance in gynecologic brachtherapy demonstrate the ability to optimize tumor coverage and reduce radiation dose to normal tissues, resulting in improved outcomes for patients.

In this article we describe a methodology to aid applicator placement and treatment planning for 3 Tesla (3T) MR-guided brachytherapy that was developed specifically for gynecologic cancers. This has been used in 18 cases to date in the Advanced Multimodality Image Guided Operating suite at Brigham and Women's Hospital. It is comprised of state of the art methods for MR imaging, image analysis, and treatment planning. An MR sequence using 3D-balanced steady state free precession in a 3T MR scan was identified as the best sequence for catheter identification with ballooning artifact at the tip. 3D treatment planning was performed using MR images. Item in development include a software module designed to support virtual needle trajectory planning that includes probabilistic bias correction, graph based segmentation, and image registration algorithms. The results demonstrate that 3T MR has a role in gynecologic brachytherapy. These novel developments improve targeted treatment to the tumor while sparing the normal tissues.

## 1. INTRODUCTION

Gynecologic malignancies, which include cervical, endometrial, ovarian, vaginal and vulvar cancers, cause significant mortality in women worldwide. In the United States, the number of gynecologic cancers has been increasing in recent years, while the death rate has remained relatively steady at about 35% of incidence [1]. The standard-of-care treatment for many primary and recurrent gynecologic cancers consists of external-beam radiation followed by brachytherapy [2]. In contrast to external-beam radiation treatment, in which a linear accelerator aims radiation beams at the pelvis from outside the body, in

Address requests to: Akila N. Viswanathan, MD, MPH Department of Radiation Oncology Brigham and Women's Hospital 75 Francis St, L2 Boston, MA 02115 Tel : (617) 732-6331 Fax: (617) 732-7347 aviswanathan@lroc.harvard.edu.

**Conflict of Interest notification:** The authors have no conflicts of interest to report.

brachytherapy, radioactive sources that deliver very high doses of radiation are placed directly inside the cancerous tissue using intracavitary or interstitial applicators. The focus of our work is on gynecologic brachytherapy, a procedure in which the use of magnetic resonance (MR) guidance holds great promise.

The use of imaging to assist with gynecologic brachytherapy treatment planning and dose delivery has evolved from 2D plain x-ray to 3D, including computed tomography (CT) and magnetic resonance imaging (MRI) [3]. With plain x-ray imaging, the exact dose administered to the tumor is unknown. Therefore, in 2D imaging, proper positioning of the applicator is critical so that dose is delivered in as symmetric a fashion as possible; proper applicator position has been shown to impact disease-free survival [4]. With 3D imaging, in addition to noting applicator position, the tumor may be visualized and contoured, permitting accurate tailoring of the radiation dose.

Although MRI is used routinely in the diagnosis of cervical cancer due to its increased sensitivity compared to CT [5], the implementation of MR in gynecologic brachytherapy planning is gradually increasing [6, 7]. Our previous studies using MR imaging in gynecologic brachytherapy demonstrate the feasibility of guidance for applicator placement using a low-field 0.5T open-configuration scanner [8, 9].

In this article we present a combined diagnostic, guidance and treatment planning approach with imaging from a single high field 3-Tesla (3T) closed-bore MR scanner. We describe our developments to aid applicator placement and treatment planning for 3T MR-guided brachytherapy specifically for gynecologic cancers, and report on our experience implementing this system in the Advanced Multimodality Image Guided Operating (AMIGO) suite at Brigham and Women's Hospital.

## 2. MATERIALS AND METHODS

### 2.1 the Advanced Multimodality Image Guided Operating (AMIGO) Suite

AMIGO was launched in 2011 as a multimodal successor to the original 0.5T Signa SP (GE Healthcare) magnetic resonance therapy (MRT) unit at Brigham and Women's Hospital, in which interstitial gynecologic brachytherapy was performed from 2002 to 2006. AMIGO is an integrated operating suite in which multidisciplinary patient treatment may be guided by x-ray, ultrasound, intra-operative 3T MRI, and/or positron emission tomography/computed tomography (PET/CT) (Figure 1). A layout of the AMIGO unit is shown in Figure 2. The goal of AMIGO is to better define anatomic tumor boundaries and the relationship of tumor to normal tissues using available imaging, and to integrate this with more complete tumor treatment via excision, ablation or irradiation.

The MRI room of AMIGO (Figure 1A) is centered on a ceiling mounted Siemens Verio MR scanner. This is a high-field (3T) wide-bore (70-cm) MRI scanner integrated with video monitors, surgical lights, therapy-delivery equipment, an MRI-compatible anesthesia machine, and vital-signs monitor. The ceiling mounted MRI scanner can be moved out of the MR room and into the operating room (OR) (Figure 1B). With this innovation, the patient does not need to be transferred from the OR table for MR imaging. The familiar "in-out" paradigm can also be used, in which the patient is imaged and then withdrawn from the bore of the scanner for intervention. In some procedures, the physician can reach into the scanner's bore to access the patient. These features enable flexibility in workflow to tailor procedures to the needs of the patient.

The OR in AMIGO is located between the MR and the PET/CT rooms (Figure 1C). It is outfitted with an electronically controlled patient table surrounded by imaging modalities and therapy devices, including a ceiling-integrated Brainlab (Brainlab AG, Feldkirchen, Germany) navigation system, a Siemens (Siemens AG, Erlangen, Germany) Artis Zee angiography unit, and two 3D ultrasound imaging systems - the Siemens S2000 and the BK (Analogic Corporation, Peabody, MA) Profocus Ultraview. The table top can be changed to optimize the procedure for surgery or angiography. All images and data pertinent to the procedure are collected using video-integration technology, prioritized, and then displayed on large LCD monitors at points of use in all three rooms in the suite.

## 2.2 Clinical Approach

From September, 2011 to April, 2012, 18 patients have undergone gynecologic brachytherapy in the AMIGO suite. Three categories of gynecologic procedures have been performed in the AMIGO suite: intra-cavitary procedures, used in 6 patients to date, which involves the insertion of an MR compatible tandem coupled with ovoids (T&O) or a ring (T&R); procedures involving hybrid applicators in 3 patients, that is a T&O with needle bearing ovoids (Utrecht Applicator) or T&R with needle bearing ring (Interstitial Ring or Vienna Applicator); and interstitial procedures in 9 patients, where plastic needles are inserted into a patient either through a Syed-Neblett template or freehand, sometimes in conjunction with a tandem but without ovoids or a ring. The interstitial needles consist of a plastic sharp-tipped outer catheter (ProGuide Needles, Nucletron Co, Veenendaal, Netherlands) with a tungsten alloy obturator that fits through the center of the needle for stabilization and identification on imaging (Figure 3A). Tandem-based applicators (Figure 3B and 3C) are made of plastic (Nucletron Co). All patients to date received high-dose-rate (HDR) brachytherapy. All patients treated in the AMIGO suite have been enrolled on an IRB-approved clinical trial for gynecologic brachytherapy, listed on the NIH website at www.clinicaltrials.gov.

Pre-implantation procedures follow the standard recommendations for gynecologic brachytherapy as outlined by the American Brachytherapy Society [2, 10, 11]. The choice of applicator is determined by the physician based on clinical examination at the time of diagnosis, during external beam, and the immediate pre-implantation assessment of disease extension. A clinical examination assesses the size of the tumor and cervix if present, the location of the uterus, the presence and extent of any vaginal disease, whether the tumor extends into the parametrial or uterosacral tissues, or to the pelvic sidewall, on rectovaginal examination.

## 2.3 Workflow for MR-Guided Gynecologic Brachytherapy in AMIGO

### 2.3.1. Anesthesia (OR room, Figure 1B or MR Room for interstitial only cases)
—Epidural with general anesthesia is used for all interstitial cases, and the epidural is continued throughout the inpatient hospitalization. General anesthesia alone is induced for all intracavitary patients treated as outpatients.

### 2.3.2. Diagnostic MR (MR room, Figure 1A)—An 8-channel spine coil is placed under the patient, and a second 8-channel body matrix coil is placed on top of the patient to provide coverage of the entire pelvic area. Patients may be placed in a modified dorsal lithotomy position for imaging. A pre-implantation 3T MRI with T2 weighted, diffusion weighted and fat-suppressed images in the sagittal, axial and coronal images is performed. The T2-weighted 3D-SPACE (3D Fast Spin Echo) images for pre- and post-implantation imaging include a slice width = 1.8 mm, FOV $320 \times 208$ mm, 96 slices/slab, 1.4 averages, TR/TE = 2500/238 msec, and bandwidth = 521 Hz/pixel. Patients receiving tandem intracavitary only insertions may be awake for this MRI and then will have general anesthesia induced in the main OR or in the MR room after completing this initial MR. All interstitial cases will have the epidural and general anesthesia prior to this initial MR.

### 2.3.3. Applicator insertion (OR room, tandem insertion required with ultrasound guidance; MR room, interstitial only cases)—A sterile speculum is inserted into the vagina to allow adequate visualization of the cervical os, the vagina and the tumor. For patients with an intact cervix, cervical dilators are inserted serially into the cervical os, often with ultrasound guidance, and a tandem is inserted into the uterus. Due to the use of ultrasound and the need to raise the legs in lithotomy position, tandem insertion for cases requiring an ultrasound is performed in the OR room then the patient is transferred to the MR room. For intracavitary cases, the ring or ovoids are positioned over the tandem and pushed against the surface of the cervix.

For template-based interstitial patients, the entire procedure is conducted in the MR room. A single suture is threaded into the vaginal apex. This suture is used to retract the apex in as far inferiorly as possible. A vaginal obturator is inserted over the tandem in cases with an intact uterus or over the suture and against the vaginal apex in cases with no tandem. Prior to placement, the holes in the template are filled with sterile surgical lubricant to enhance visibility in subsequent MR scans. The template is sutured to the patient's perineal skin in the four corners of the template.

### 2.3.4. Guidance MR (MR Room, Figure 1A)—a. An initial localizer T2 sagittal, axial and coronal MR is performed with the applicator in situ. Software is being tested that performs image analysis. allows visualization of the applicator relative to the anatomy, and plans the placement of virtual needles when applicable.

b. For interstitial cases, 24-29 cm-long ProGuide (Nucletron Co) hollow plastic catheters with metal central obturators are inserted in the holes surrounding the plastic intra-vaginal obturator. After insertion of the first needles, a 3D-balanced steady state free precession (bSSFP) sagittal image (Figure 4) is obtained to localize the most superior aspect of the catheter. Adjustments to the most superior position are made to ensure no inadvertent insertion of a catheter into the bowel and adequate superior tumor coverage. Additional interstitial catheters are inserted, with the insertion depths intended to completely surround the visualized areas of the tumor and avoid normal tissues. A series of quick monitoring > 1 minute T2 MR scans are acquired periodically as the needles are inserted and advanced to the tumor region. Software is used to visualize the inserted needles and plan the next ones.

c. Confirmation MR scans consisting of T2 weighted sagittal, axial and coronal imaging with slice thicknesses of 1.6 mm are acquired at the end of applicator/needle placement. For intracavitary tandem and ring/ovoid procedures, a 1 mm slice thickness 3D Axial SPACE scan can be performed.

### 2.3.5. Radiation treatment planning and administration (Radiation Oncology Clinic)—MR images are transferred via the hospital network into the treatment planning system (Oncentra Brachy, Nucletron Co.) where contouring of tumor and surrounding bladder, rectum and sigmoid, and dosimetric calculations are performed. For interstitial cases performed through February, 2012, a confirmatory CT scan was obtained to ensure accurate catheter identification. As of March, 2012, planning directly from the MR has been implemented in selected cases. After appropriate safety and quality assurance checks [2, 10-12], radiation therapy is delivered based on the treatment plan with high-dose-rate (HDR) brachytherapy in a room designated for radiation treatment with appropriate shielding. Patients undergoing intracavitary treatment are discharged at the end of the day, while those undergoing interstitial treatment are admitted as an inpatient with epidural anesthesia to continue over the next 5-6 days while they receive HDR fractions twice a day throughout the week.

# 3. RESULTS

## 3.1 Catheter Identification

A pressing problem in MR-guided interstitial brachytherapy is difficulty with precisely locating the catheters during the insertion process. Therefore, we sought to identify an MR sequence that could be used to reproducibly identify each catheter after insertion. In order to visualize and distinguish the catheters from each other, multiple contrast MR sequences (3D-bSSFP, 3D-fast spin echo (3D-FSE), 3D T1-weighted gradient echo (3D-GRE) were acquired. After insertion of the first catheter, an attempt is made to identify the tip of the catheter after each of these sequences.

When comparing the 3D-FSE (Figure 4a, 4c), 3D-GRE, and 3D-bSSFP series, the 3D-bSSFP sequence provides the best visual information for catheter identifications with ballooning artifact at the tip on a sagittal image (Figure 4b). On an axial scan, 3D-bSSFP produces needle-artifact that results in a cross shape at the center of the catheter (Figure 4d). After a series of modifications to reduce the size of the balloon at the tip by changing the 3D-bSSFP bandwidth and repetition-time (TR), an optimized sequence was incorporated into the workflow with TR/TE/$\theta$=4.6 ms/2.3 ms/30$^0$, bandwidth= 600 Hz/pixel, 0.9×0.9×1.6 mm$^3$ resolution, 160 mm superior-inferior coverage, 1.2 minute/acquisition. The most accurate identification was confirmed with this optimized 3D-bSSFPP series. In contrast, the 3D T1-weighted GRE (TR/TE/$\theta$=20 ms/7-9 ms/30$^0$ , bandwidth= 800 Hz/pixel, 0.9×0.9×1.2 mm$^3$ resolution, 160mm Superior-Inferior coverage, 3 minute/acquisition) signal-to-noise ratio was lower, and an equivalent-resolution 3D-FSE (Siemens SPACE TR/TE=2500ms/ 140ms, bandwidth=800 Hz/pixel, 0.9×0.9×1.2mm$^3$ resolution, 160 mm superior-inferior coverage, 2X GRAPPA acceleration) required 5-minute scans in contrast to the 1.2 minute 3D-bSSFP. Accurate identification of the catheter tip during the insertion process allows proper placement and ensures that catheters are not unintentionally inserted into the normal-tissue structures.

## 3.2 Treatment Planning

We define treatment planning as the process of designing the delivery of radiation through the implant and the calculation of the dose given to the patient. This definition is not inclusive: the concept of treatment planning can be expanded to embrace the choice of applicator and the specifics of the implantation (e.g., where the interstitial needles are positioned). The separation of the implantation planning from the treatment planning was dictated by the different role that the MR plays in these two phases of treatment management. Figure 5 shows the typical workflow of the planning process from the end of the implantation to the optimized plan.

The concept of 3D treatment planning refers to the digitization of the location of the implant, optimization of the dose distribution, and calculation of dose volume histogram (DVH) to the OARs. 3D dose optimization is not widely practiced yet [3], despite the wide availability of 3D imaging to guide and evaluate implantations. 3D planning based on MR is fairly recent, with some clinics reporting good results when an MR-based high-risk clinical target volume (HR-CTV) was used for optimization [14-16].

### 3.2.1 Post-Implantation MR—As described in workflow section 2.3.4c, all patients receive an axial, sagittal and coronal T2 MR scan with a slice thickness of less than 3 mm immediately after applicator insertion.

### 3.2.2 Additional Imaging and Registration—Patients treated with intracavitary applicators do not undergo additional imaging. Patients treated with interstitial applications

from September, 2011 to March, 2012 all underwent a CT scan with a 1.25-mm slice thickness in a dedicated CT scanner in the Radiation Oncology brachytherapy suite to aid with proper localization of the implant. Since March, 2012, interstitial patients have undergone MR-based treatment planning using an axial T2 MR with a slice thickness of 1.6 mm registered with the b-SSFP images described in section 3.1. For patients undergoing CT simulation after MR-guided insertion, copper radioopaque markers are inserted in the needles and commercially available markers are inserted into the T&O or T&R for ease of identification on CT. CT and MR images are registered based on the fusion of the applicator. The applicator is used as a proxy for the HR-CTV soft tissue, which is not easily distinguishable on CT.

### 3.2.3 Tumor and Normal Tissue Contouring

Clinical contouring (manual segmentation) is performed on the available image sets. For intracavitary patients, contouring is purely MR-based. For other procedures, MR is used for all tumor contouring, but normal tissue structure contouring is performed on the CT/MR fusion with comparison of the normal tissue outlines. MR may be used for OAR contouring in when CT cannot distinguish OAR borders, for example when segments of bowel rest close to the cervix. For interstitial cases, CT may be preferable to ensure quick and accurate digitization of the needles. HR-CTV is contoured on MR following the guidelines of The Groupe Européen de Curiethérapie and the European Society for Therapeutic Radiology and Oncology (GEC-ESTRO) [17, 18]. An intermediate-risk vaginal structure is contoured if this is required.

MR scanning in gynecologic applications allows for easy identification of the HR-CTV. This translates into three advantages as compared with CT: during implantation, the applicator and the needles can be located in closer proximity to their optimal position in relation to the tumor for radiation delivery; contouring is faster as a second scan is not required; and contouring is more accurate, as the tumor may be visualized resulting in an MR-based high-risk clinical target volume (HR-CTV) [15].

### 3.2.4 Applicator reconstruction

Applicator reconstruction refers to the digitization of all possible locations of the source in the applicator. Due to the sharp dose gradients present in brachytherapy, applicator reconstruction is a critical part of treatment planning, and has the potential for introducing large uncertainties into the dose delivery [19]. Two methodologies for applicator reconstruction are available: manual and model-based.

Manual reconstruction requires the manual digitization of the path available to the source along a catheter. We manually digitize all needles on the CT images. CT scans are performed with dummy markers inserted inside the needle. These copper markers are radio-opaque along the path, and have a 1-cm radio-transparent section before the last 0.5-cm radio-opaque tip. This configuration allows for easy digitization of the needle and easy identification of the catheter tip.

Model-based reconstruction allows the planner to superimpose a model on the applicator that is visible on the scan. As long as the shape of the applicator contains enough details for correct registration of the model to the image, it is not necessary to visualize in detail the path available to the source inside the applicator. We use model-based reconstruction for T&R and T&O applicators.

All our patients are implanted under general anesthesia, so it is a priority to minimize both anesthesia time and X-ray dose. Avoiding CT imaging after MR is therefore advantageous. We performed 3 successful tests of T&O/T&R plastic applicator reconstruction on MR in patients treated in AMIGO, as compared with reconstruction on a CT acquired immediately after implantation. We concluded that reconstruction of these applicators was possible directly on the T2-weighted axial MR, aided by fusion with the concomitant T2 sagittal MR. This result is in line with the literature [20, 21]. We observed a high degree of uncertainty in needle reconstruction on T2 MR prior to development and implementation of the bSSFP series. Other clinics have reported good results with direct MR reconstruction of applicators including needles [22]. These results typically apply to needles anchored to ovoids, with an average insertion depth of 2.5 cm inside the patient from their anchorage on the applicator, and a maximum of 10 needles. For interstitial procedures performed recently in our clinic, a median of 15 needles were used with a depth typically exceeding 10 cm. The visibility of the needles degraded with the number of needles and with the depth of insertion as visualized on T2 images, making the needles in interstitial procedures particularly difficult to reconstruct on MR. The use of an axial T2 MR with a slice thickness of 1.6 mm registered with the b-SSFP images described in section 3.1 represents a significant advance in MR-based needle reconstruction. The b-SSFP MR enhances the visibility of the needle tip locations, and can be used to better distinguish needles that are close together. Nevertheless, MR-based needle reconstruction is more time-consuming than CT-based needle reconstruction, and our clinical practice is to obtain CT scans for the majority of patients in whom needles have been implanted, as this has been the fastest way to ensure accurate digitization of the needles.

**3.2.5 Treatment Planning—**3D planning is an iterative process. The starting point of a T&O/T&R treatment plan is a preliminary 2D plan normalized to Point A, with a standard loading pattern of the applicator that provides a dose distribution commonly referred to as pear-shaped (Figure 6a). For hybrid applicators, the 2D plan is modified so that needles are loaded from the tip down to 1 cm away from the needle-bearing ovoid or ring, and dwell times are adjusted so that overall loading time is distributed 85% in the intra-cavitary applicator and 15% in the needles (Figure 6b). For interstitial implants, the preliminary plan consists of a uniform loading of all needles and Point A is not defined (Figure 7a).

These standard plans are not in general satisfactory. In the example in Figures 6a, 6b, and 7a, coverage of the HR-CTV is inadequate. In other cases, prescription dose might be covering areas past the HR-CTV into the OARs.

3D optimization is the process of adjusting the standard plan to a specific case based on the segmentation of the CT or MR scan. The dose distribution is optimized by manually adjusting the dwell times in specific locations to maximize coverage of HR-CTV, while dose to the OARs is minimized. An effort is made to minimize deviations from the standard pear-shape distribution. Needles, if present, are loaded lightly, although no policy is in place on a maximum acceptable loading of a needle in hybrid applicators (Figure 6c). For interstitial implants, a mix of manual and graphic optimization is at times performed, depending on the size of the implant and the planning time constraint (Figure 7b). A final review of dwell times is always manually performed to minimize variations of dwell times among dwell locations.

The attending radiation oncologist and the planner make final adjustments by iteratively adjusting dwell times and checking the distribution of isodose lines and DVH. The final plan is a compromise between maximization of HR-CTV V100, HR-CTV D90 exceeding 100% of prescription dose, and minimization of D2cc for rectum, bladder and sigmoid. Our methodology and our results are in line with the 3D optimization experience reported by other clinics [14, 23]

In summary, MR-based 3D planning increases coverage to the HR-CTV and reduces the dose to the OARs. T&O and T&R applications can be planned directly on MR. If needles are inserted, a T2 MR with 1.6 mm slice thickness and a b-SSFP MR may be sufficient for needle identification, and a CT scan may be acquired and fused onto the MR scan to help with reconstruction and quicken the digitization process. HR-CTV contouring on the T2 MR in these cases is more precise than CT-only planning.

### 3.3 Future Directions

**3.3.1 Interleaving Dose Planning and Needle Guidance**—Recognizing that less than ideal needle placements can often be compensated for by the degrees of freedom available in the needle after loading step, early work from our group demonstrated a method for incorporating the observed needle positions from a 0.5T intraoperative MR scanner into the treatment plan for prostate brachytherapy using permanent implants [24]. We will build upon that method and interleave the needle guidance and treatment planning steps in partnership with commercial vendors of treatment planning software.

**3.3.2 Needle Trajectory Planning**—An image-guided navigation system aids in the precise placement of needles, catheters, and other instruments at prescribed locations in the patient's body, as shown by pre- or intra-operative images, and then monitors their trajectory through the course of the intervention [13]. Commercially successful navigation products are available for brain-tumor resection (e.g., Brainlab Vector Vision, Brainlab AG, Germany), cardiac EP ablation (CarTo3, Biosense Webster Inc., Daimond Bard, CA; ensite-NavX velocity, St. Jude Medical, St Paul, MN), spine surgery (e.g., Medtronic Stealth Station, Medtronic Corporation, Louisville, CO) and, more recently, abdominal ablation (e.g., Percunav, Philips Corporation, Best, Netherlands) and lung bronchoscopy (e.g., SPiN Drive, Veran Medical Technologies Inc, St. Louis, CO). In the pelvis, there are indications of progress in "permanent" seed implant prostate brachytherapy navigation products (Radvision, Acoustic MedSystems Inc., Savoy, IL), but none that we are aware of for gynecologic brachytherapy. We have designed a navigation software application, *iGyne*, specifically for gynecologic brachytherapy procedures, and describe its needle-trajectory planning component. The needle-trajectory planning method focuses on Syed-Neblett template-based interstitial HDR brachytherapy, although extensions to cover different applicators are in development.

**3.3.3 *iGyne* Software Application**—The *iGyne* software, in development, may be utilized as part of the process in section 2.3.4 of the MR-guided brachytherapy workflow. After the template is sutured to the patient's perineum, the iGyne software may be used to perform a rigid registration of a CAD model of the template to its image in the MRI scan, using corresponding markers that are clearly visible in the MRI. Figure 8A presents the basic principle for fitting a Syed-Neblett template to a patient dataset using *iGyne*. For template fitting, the user selects three markers in the patient's scan (three squares in the upper left window) which correspond to three positions surrounded in red in the CAD model of the template (upper right window) and three positions marked by red circles in the planning sheet (lower right window). Figures 8 further illustrate the principle of needle planning with the *iGyne* software prototype. Figure 8B shows the selection of a specific interstitial needle (Ba, red circles in the screenshot) after the model of the gynecologic brachytherapy template has been added to the initial patient scan. As shown on the left side of the prototype interface, the depth and trajectory of each interstitial needle can be planned individually by defining parameters such as needle length and needle depth. Figure 8C shows virtual placement of several interstitial needles (purple) with different lengths and depths with the settings in the menu on the right side of the software prototype. Figure 8D presents how a specific needle can be selected and how the needle path can be displayed under *iGyne*. The selected needle (Ad) is shown as a white line in the upper right window. The user can now move a multiplanar reconstruction (MPR) along the needle path that is visualized in the 3D view (upper right window) and the MPR itself is visualized as a 2D slice in the lower left window. In the presented screenshot, the MPR at the position of the arrow (tip of red arrow in the upper right image) is displayed in the lower left window. In addition, the needle (white) is surrounded by a red circle in the MPR (lower left window). Though this may be helpful for pre-planning, it will be necessary to modify the program to allow for needle deflection. Therefore, *iGyne* has not yet been used to facilitate needle-trajectory planning as the physician inserts multiple needles but has potential for future implementation.

### 3.3.4 Image Analysis: Probabilistic Bias Correction of MRI Spatial Signal Inhomogeneity—We will use a probabilistic bias correction method to explicitly model and correct for the corrupting field in MRI images that is due to the spatial inhomogeneities in the radiofrequency (rf) field produced by the surface coils that are placed above and below the patient. Wells et al. introduced this Bayesian method [25] which has been further developed [26-28]. This method is based on the Expectation-Maximization algorithm, an estimation method that is used to find the parameters of a model that maximizes the likelihood of the data when some of the data are observable and some are hidden or unobserved. The earliest thorough treatment of the applications of Expectation-Maximization appeared in a report by Dempster et al. [29]. Starting from an initial guess, Expectation-Maximization re-estimates the model parameters in an iterative manner. Each iteration consists of two steps: an Expectation or E step and a Maximization or M step. In the E-step, the probability distribution for the hidden variables is computed, given the known values for the observed variables and the current estimate of the model parameters. In the M-step, the parameters that maximize the likelihood of the data are reestimated assuming that the probability distribution computed for the hidden variables in the E-step is correct. It can be shown that the likelihood does not get worse with each iteration of the EM algorithm.

### 3.3.5 Image Analysis: Segmentation of Organs at Risk—As a first step towards an overall segmentation of anatomy structures for gynecological brachytherapy, though many alternative approaches exist, we segmented the bladder with the Nugget-Cut approach[30] [31]. Figure 9 presents the overall segmentation principle.

## 4. Discussion

In this overview, we describe our ongoing work in gynecologic brachytherapy in the recently opened AMIGO suite at BWH. We describe the integration of the available imaging modalities as used to date, with future work in PET scheduled to begin in the near future.

Prior to the development of the AMIGO suite, in a clinical trial from 2002-2006, we demonstrated the benefit of intra-procedural MR images in 25 patients both in identification of parametrial tissue involvement [9], and in optimized radiation delivery by performing interstitial needle placement in an open configuration 0.5 Tesla MR scanner (Signa SP, GE Healthcare, Waukesha, WI at that time). In the years between 2005 and 2011, two dedicated CT-based brachytherapy suites were constructed in the Department of Radiation Oncology. In these suites, intra-procedural CT scans, in contrast to the use of plain film x-ray only, increased the precision of placement of intracavitary applicators for gynecologic brachytherapy. In parallel, the discovery and optimization of MR image sequences improved in diagnostic MR imaging, allowing translation of these MR findings to gynecologic brachytherapy [32, 33].

Over the past four months, we have assessed several technologies, including accurate interstitial catheter identification, MR bias correction, bladder segmentation, and developed a novel software program to assist with accurate catheter placement. Ongoing areas of focus are: 1) active tracking of the tips of interstitial needles in the MR scanner; 2) fabrication of customized needles that generate a more distinct artifact in MRI and, in particular, a unique artifact at the needle tip; 3) MR pulse sequences to enhance needle artifacts; 4) a model-to-image registration method which aligns the geometric model of the template to its MRI image; 5) improved image-based needle detection algorithms to precisely demarcate the inserted needles; 6) extension of the Nugget-Cut algorithm to segment additional organs at risk; and 7) development of real-time iterative treatment planning modules available in the AMIGO suite. We expect to be able to make rapid strides in each of these areas. Some of these advances may continue by translating advances from prostate-cancer brachytherapy [34-36], MRI physics [37], active MRI tracking of cardiac catheters [38, 39], and image processing for brain tumors and prostate cancer [40, 41]. We also anticipate developing unique and novel strategies for gynecologic brachytherapy that may be translated to other fields, and encourage colleagues who are considering establishing MR-guided gynecologic brachytherapy.

## 5. Conclusion

MR image-guidance for brachytherapy of gynecologic malignancies permits accurate placement of the applicator and analysis of tumor location in relation to surrounding normal tissues. Advances in areas of MR bias correction, image-applicator identification, tracking, organ segmentation, and radiation treatment planning provide unique opportunities to improve the process, increase speed and ideally improve outcomes for patients.

## Acknowledgments

The authors would like to acknowledge the support of the AMIGO clinical staff in enabling this clinical study, Sam Song, PhD for generating the CAD models of gynecologic devices and Junichi Tokuda, PhD for the AMIGO photo. We thank Prof. Dr. Horst K. Hahn and Fraunhofer MeVis in Bremen, Germany, for their support by providing an academic license for MeVisLab software. We wish to thank Barbara Silver for proofreading the manuscript. The work was supported by NIH grant P41EB015898, P41RR019703, R03EB013792, and U54EB005149. Dr. Viswanathan has received support from NIH grant K07CA117979.

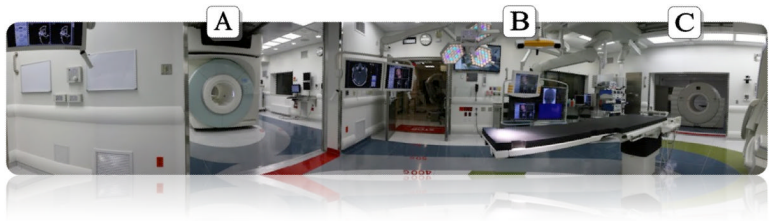

**Figure 1.**
AMIGO Suite with A) MR room, B) OR, and C) PET/CT Scanner.

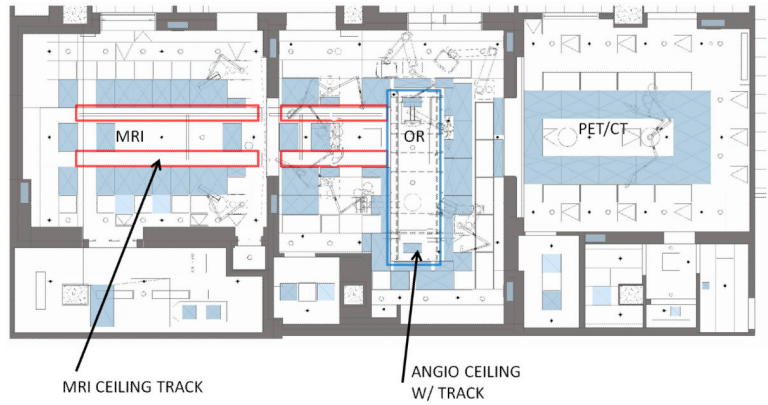

**Figure 2.**
Lay-out of the AMIGO floor plan with the MRI room and the MRI ceiling track (left), the OR and the angio ceiling (middle) and the PET/CT Scanner room (right) corresponding to Figure 1.

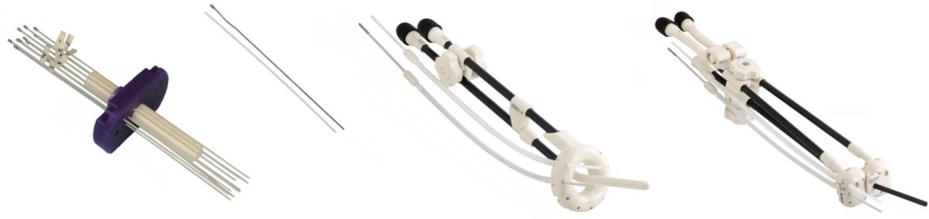

**Figure 3.**
Applicators used in gynecologic brachytherapy include A) Interstitial catheters with central tungsten alloy obturators placed through a central vaginal obturator (white) and through a disposable template; B) tandem and ovoids; and C) tandem and ring; B and C may be used with or without interstitial catheters.

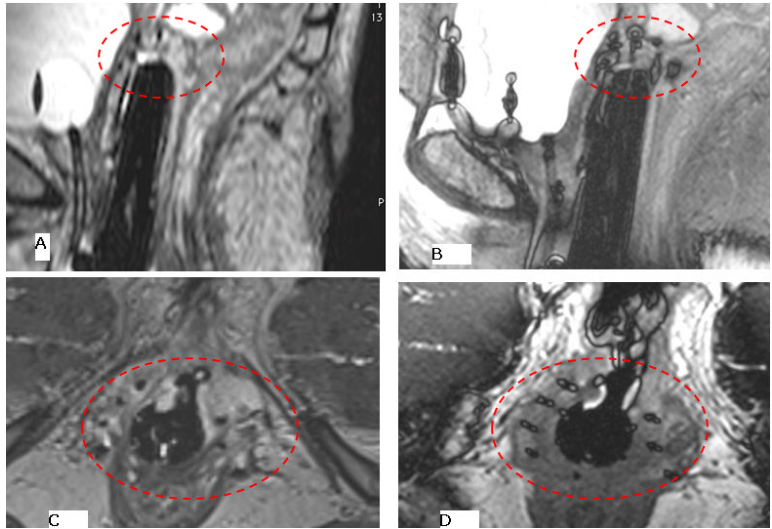

**Figure 4.**
Ballooning created at catheter tip using the 3D-bSSFP sequence. A) a sagittal image using using fat suppressed 3D-FSE 1.2 mm slice width, taken over approximately 5 minutes shows the difficulty in identifying catheter tips, whereas the B) 3D fat suppressed balanced SSFP, 1.6 mm slice width, over approximately 1.2 minutes allows rapid identification of the catheter tip. This determines the deepest point of insertion, in order to avoid bowel insertion, and covers the length of the tumor. All subsequent needles are inserted to a similar depth based on tumor location. Similar results are seen on axial images C) 3D-FSE and D) 3D balanced SSFP, where instead of the balloon configuration seen on the sagittal image, a cross centered on each catheter can be visualized.

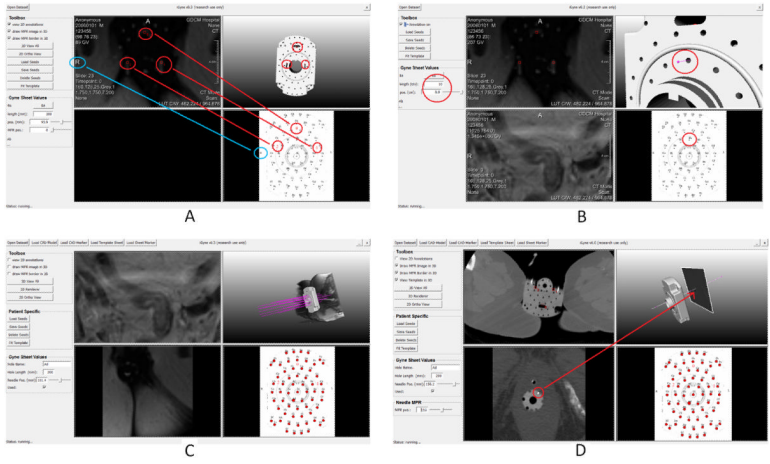

**Figure 5.**
Workflow of the treatment planning process for AMIGO procedures.

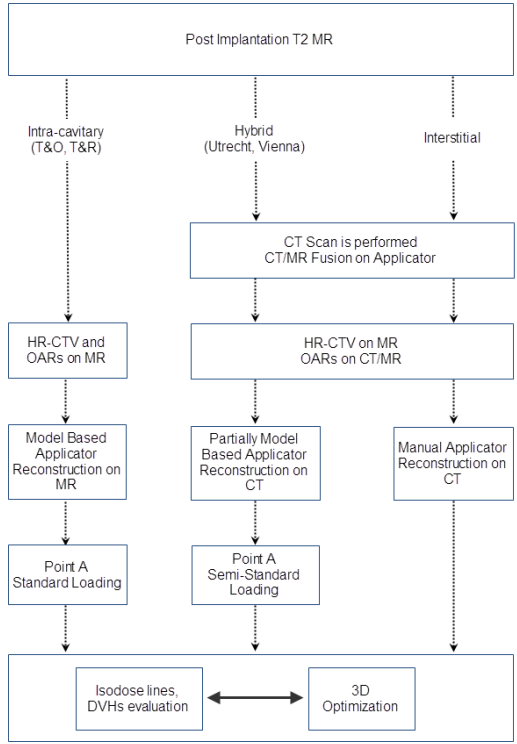

**Figure 6.**
Tandem and ring (T&R) with interstitial needles. A) The preliminary plan consists of a standard loading to the T&R, resulting to a pear-shaped distribution with prescription dose (100% isodose line in yellow) to the A points. B) Interstitial needles are uniformly loaded, with dwell times amounting to 15% of the total dwell loading time. C) 3D optimization of T&R and interstitial needles results in increased HR-CTV coverage and sparing the OARs.

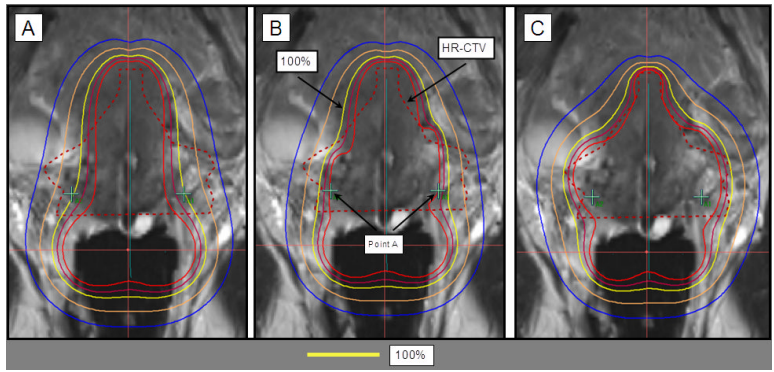

**Figure 7.**
Interstitial treatment planning with the 100% isodose line in yellow. A) The preliminary plan consists of a uniform loading of all dwell positions. B) 3D optimization of the dwell loading results in increased coverage to the HR-CTV and sparing the OARs.

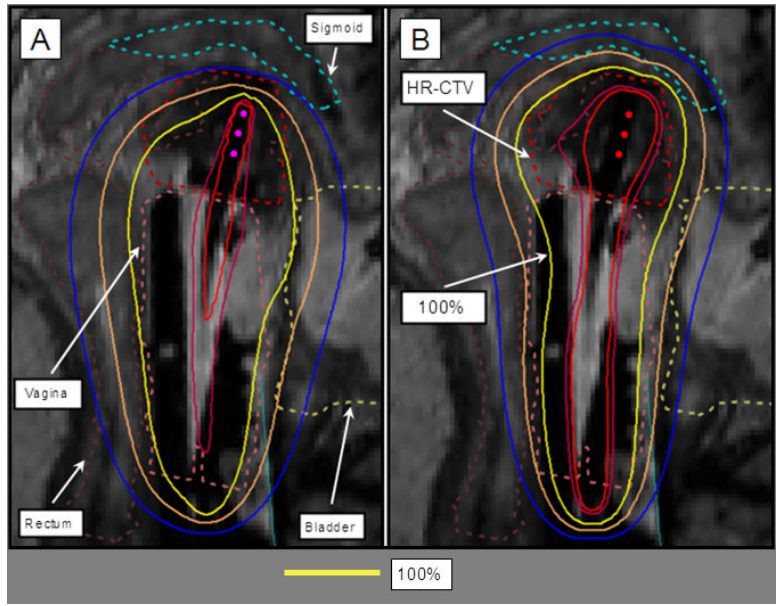

**Figure 8.**
Principle for fitting a gynecological template for brachytherapy with the initial MRI image. The three red circles indicate corresponding needle holes in the template and the MRI image. The fitting is realized via a rigid transformation between these corresponding point sets. The blue circles are used to ensure that the left and right sides of the patient and the template are matched correctly (A). Virtual fitted gynecological brachytherapy template and selection of a specific interstitial needle (Ba, red circles in the screenshot). As shown on the left side of the prototype interface, individual needle insertion can be planned by defining parameters such as the needle length and depth (B). Virtual placement of several interstitial needles (purple) with different lengths and depths as shown in the settings in the menu in the left column. This allows the radiation oncologist to plan the placement of needles (C).Needle (white line in the upper right window) that has been selected for visualization of multiplanar reconstructions (MPR) along the needle path (lower left window). The MPR at the position of the arrow (tip of red arrow in the upper right image) is displayed in the lower left window as a 2D slice. In the MPR of the lower left window the needle cross section (white) is surrounded by a red circle (D).

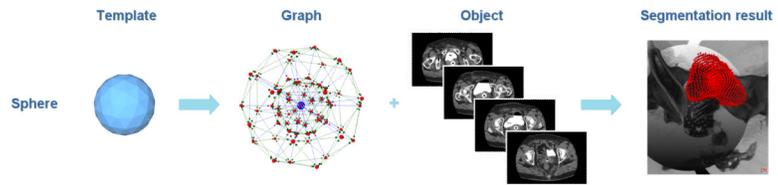

**Figure 9.**
Segmentation principle of the bladder with the Nugget-Cut approach: a spherical template (left image) is used to construct a directed graph (second image from the left). The graph is fused with the CT dataset of a patient (third image from the left) and the s-t-cut provides the segmented bladder (red, rightmost image).